\documentclass[epj,twocolumn]{webofc}
\usepackage[varg]{txfonts}   % Web of Conferences font
\woctitle{ISVHECRI 2016}
\usepackage{psfrag}

\newcommand{\rr}{\mathbf{r}}

\begin{document}
\title{Particle Spectra and Mass Composition in the Ultra-High Energy Region in the Framework of the Galactic Origin of Cosmic Rays}

\author{
\firstname{A.A.} \lastname{Lagutin}\inst{1}\fnsep\thanks{\email{lagutin@theory.asu.ru}}
\and
\firstname{N.V.} \lastname{Volkov}\inst{1}
\and
\firstname{A.G.} \lastname{Tyumentsev}\inst{1}
\and
\firstname{R.I.} \lastname{Raikin}\inst{1}
}

\institute{Altai State University, Radiophysics and Theoretical Physics Department, 656049, 61 Lenin ave, Barnaul, Russia 
          }

\abstract{%
The possibility for a self-consistent description of all the basic features of the observed cosmic ray spectra and primary composition variations in the energy range of $10^{15}\div 10^{20}$~eV within the Galactic origin scenario is examined.

We assume the existence of Galactic sources that accelerate particles up to $\sim 3\cdot 10^{18}Z$~eV and take into account a highly inhomogeneous (fractal-like) distribution of matter and magnetic fields in the Galaxy that leads to extremely large free paths of particles (``L\'{e}vy flights''), along with an  overwhelming contribution to the cosmic ray fluxes observed above $\sim 10^{18}$~eV from particles reaching the Solar System without scattering. Our scenario was refined on the basis of recent experimental results on primary mass composition. Model predictions, which could be verified with the improved high-precision measurements in the nearest future are discussed.}

\maketitle

\section{Introduction}\label{intro}

Physical mechanisms governing the features of energy spectra of cosmic rays (CRs) and their composition variations in the energy range that spans from PeV to tens of EeV are still under debate.
The so-called standard scenario assumes that the Galactic component of CRs accelerated by  supernova remnants extends up to energies of a few  $10^{17}$~eV with gradually heavier composition due to the rigidity-dependent maximum acceleration energy of nuclei. Then Galactic CRs merge into a light and flatter extragalactic component, which terminates at $\sim 5\cdot 10^{19}$~eV due to the GZK effect.

Recent measurements by several new generation experiments and also the results of some new methods of experimental data analysis~\cite{Apel-a:2013,Buitink:2016,Settimo:2016,Aab:2016,Abbasi:2015} have severely challenged the standard scenario. In particular, an ankle-like feature in the light CR component measured by KASCADE-Grande at $E>10^{17}$~eV~\cite{Apel-a:2013} together with  evidence for a light-mass fraction (protons and helium) of about 80 percent at $10^{17}\div 10^{17.5}$~eV~\cite{Buitink:2016} cause difficulties for an explanation in the framework of the standard model without strong additional assumptions. At ultra-high energies the chemical composition is still ambiguous. Mixed composition including nuclei heavier than helium is reported by the  Pierre Auger Observatory (PAO) on the basis of various improved methods and techniques~\cite{Settimo:2016,Aab:2016}, while the Telescope Array (TA) interpretes their data as being compatible with a largely protonic composition~\cite{Abbasi:2015}. Thus, the non-GZK origin of the energy spectrum suppression at the highest energies (e.g. the acceleration limits in Galactic cosmic ray sources above $10^{18}$~eV) can not be excluded.

In our papers~\cite{Lagutin:2005ijmp,Lagutin:2015,Volkov:2015} it was shown that basic structures in the all particle CR spectrum (knee, ankle and cut-off) can be reproduced under the assumption that  essentially all CRs, including those at the highest energies, originate from  Galactic sources, which can accelerate particles up to the cut-off energy. The highly inhomogeneous (fractal-like) distribution of matter and magnetic fields in the Galaxy that leads to extremely large free paths of particles (so called ``L\'{e}vy flights''), along with large contribution of non-scattered particles to cosmic ray fluxes above $\sim 10^{18}$~eV is the main element of our model~-- the anomalous diffusion model. At the present stage the new more precise data, mainly the spectrum features measured separately for light and heavy components along with the improved composition estimations, provide an opportunity for refinements and updates of the model.

In the present paper we revise the possibility of the self-consistent description of the CRs energy spectra and composition results within the Galactic cosmic ray origin scenario based on the anomalous diffusion model and discuss the crucial model predictions, which could be verified with  improved high-precision measurements in the near future. 

%In the following section we give an overview of the anomalous diffusion model, discuss %our injection exponent retrieval technique and the results obtained in this model. In %section 3 we present a new approach to describe the
%cosmic rays propagation in non-homogeneous ISM, based on the normal diffusion %equation. Our findings and
%conclusions are given in section 4.

\section{Key elements of the proposed scenario}

The following basic principles and assumptions were initially included in the proposed scenario.
\begin{enumerate}

\item The existence of Galactic sources that accelerate particles with  atomic charge $Z$ up to \mbox{$E_0\sim 4\cdot 10^{18}Z$~eV}. Consequently, the observed suppression of the primary CR spectrum at $\sim 5\cdot 10^{19}$~eV is not an extragalactic GZK feature; it reflects the acceleration limit of Galactic CR sources.

\item The highly inhomogeneous (fractal-like) distribution of matter and magnetic fields in the Galaxy that leads to the anomalous diffusion of CRs manifested, in particular, by abnormally large free paths of particles (so-called ``L\'{e}vy flights'') with a power-law distribution.

\item An overwhelming contribution to  cosmic ray fluxes observed above $\sim 10^{18}$~eV from particles reaching the Solar system without scattering.
\end{enumerate}

Note, that the main argument  contrary to any model assuming the Galactic origin of ultra-high energy cosmic rays is that particles originating from Galactic sources could hardly be as isotropic as presently observed. However, in recent works~\cite{Kumar:2014,Eichler:2016} it was shown that the isotropy can be understood in the context of Galactic production if a sufficiently careful treatment of CR propagation in the Galaxy is undertaken.

The spatial distribution of sources also suggests the separation of the observed CR flux into three components as follows:
$$
J(\rr,t,E) = J_G(\rr,E)+ J_L(\rr,t,E)+J_{NS}(\rr,E).
$$
Here
\begin{itemize}
\item $J_G$ is the global spectrum component determined by the multiple old ($t\geq 10^6$~yr) distant ($r\geq 1$~kpc) sources.
\item $J_L$ is the local component, i.e. the contribution of nearby ($r< 1$~kpc) young ($t< 10^6$~yr) sources. The spatial and temporal coordinates of the local sources are presented in~\cite{LagutinTymentcev:2004}.
\item $J_{NS}$ is the flux of non-scattered particles.
\end{itemize}

We assume that the non-scattered component is also formed by nearby ($r< 1$~kpc) sources, defining the spectrum in the ultra-high energy region, and provides the observed flattening of the spectrum at $E \geq 10^{18}$~eV.

The flux of non-scattered particles, $J_{NS}$, is determined by the injected flux $(S_0E^{-\gamma}\exp(-E/E_0)$, where \mbox{$E_0=4\cdot 10^{18} Z$~eV}), multiplied by a factor describing the probability of reaching the observer at a distance, $r$, from the source for a particle with a certain energy and atomic number without scattering. This factor has a power-law asymptotic with respect to $r$ (the L\'{e}vy flight probability; see the next section for the explicit form) and also suppressed at energies $E<3\cdot 10^{17}Z$~eV reflecting the fact that for the nucleus with gyroradius less than the typical Galactic inhomogeneity size this inhomogeneity would be opaque.

\section{Parameters of the cosmic ray anomalous diffusion model}

The highly inhomogeneous character of matter distribution and associated magnetic fields in the Galaxy should be adequately incorporated into the cosmic ray diffusion model. A physically reasonable way for the generalization of the normal diffusion model is to abandon the assumption about statistical homogeneity of the  distribution of matter in favour of its fractal distribution. A principle consequence of this generalization is the power-law distribution of free paths $r$ in such a medium $p(\rr,E) \propto A(E,\alpha)r^{-\alpha - 1}, r \rightarrow \infty,  0 < \alpha < 2$~--- so-called L\'{e}vy flights. Besides, the  intermittent magnetic field of the fractal-like interstellar medium (ISM) leads to a higher probability of a long stay of particles in inhomogeneities,  leading to a presence of the so-called L\'{e}vy traps. In the general case, the probability density function 
$q(t,E)$ of time $t$, during which a particle is trapped in the inhomogeneity (L\'{e}vy trap), also has a power-law behaviour: $q(t,E) \propto B(E,\beta)t^{-\beta - 1},   t \rightarrow \infty,  \beta < 1$. 

Generalization of the homogeneous normal diffusion model to the case of inhomogeneous (fractal-like) ISM, has been made for the first time in our papers~\cite{Lagutin:2000,Lagutin:2001}. Later, it was shown~\cite{Lagutin:2003,Lagutin:2003ew,LagutinTymentcev:2004,Lagutin:2009}   that an anomalous cosmic ray diffusion model allows to describe the main features of nuclei, electron and positron spectra observed in the Solar system. Particularly, in the anomalous diffusion model the key feature of the all particle energy spectrum~--- the knee at $3\cdot 10^{15}$~eV~---  appears naturally without additional assumptions. 

The equation for the density of particles with energy $E$ at the location $\rr$ and time $t$, generated in a fractal-like medium by Galactic sources with a distribution density $S(\rr,t,E)$ can be written as~\cite{Lagutin:2003,LagutinTymentcev:2004}
\begin{multline}~\label{SuperEq}
\frac{\partial N(\rr,t,E)}{\partial t}=-D(E,\alpha,\beta)\mathrm{D}_{0+}^{1-\beta}(-\Delta)^{\alpha/2} N(\rr,t,E)+\\ + S(\rr,t,E).
\end{multline}
Here $\mathrm{D}_{0+}^{1-\beta}$ denotes the Riemann-Liouville fractional derivative~\cite{Samko:1993} and $(-\Delta)^{\alpha/2}$ is the fractional Laplacian (``Riesz operator'')~\cite{Samko:1993}. The anomalous diffusion coefficient $D(E,\alpha,\beta) \sim A(E,\alpha)/B(E,\beta) = D_0(\alpha,\beta)E^{\delta}$.

The solution of Eq.~\eqref{SuperEq} for a point impulse source with a 
power-law injection spectrum and emission time $T$ $S(\rr,t,E)=S_{0} E^{-\gamma}\delta(\rr) \Theta(T-t)\Theta(t)$ ($\Theta(\tau)$ is the step function)  has the form~\cite{Lagutin:2003,LagutinTymentcev:2004}
\begin{multline}\label{eq:solanomdifeq}
N(\rr,t,E)=\frac{S_0 E^{-\gamma}}{D(E,\alpha,\beta)^{3/\alpha}} \times\\
\int\limits_{\max[0,t-T]}^{t}d\tau \tau^{-3\beta/\alpha}\Psi_3^{(\alpha,\beta)}\left(|\rr|(D(E,\alpha,\beta)\tau^{\beta})^{-1/\alpha}\right),
\end{multline}
where $\Psi_3^{(\alpha,\,\beta)}(\rho)$ is the density of the fractional stable distribution~\cite{Uchaikin:1999a,Zolotarev:1999}
\begin{equation*}
  \Psi_3^{(\alpha,\,\beta)}(\rho)=\int\limits_0^\infty{g_3^{(\alpha)}({r\tau^\beta})q_1^{(\beta,1)}(\tau)\tau^{3\beta/\alpha}d\tau}.
\vspace*{-3mm}
\end{equation*}

Using the representation $N = N_0 E^{-\eta}$ and the property $d\Psi_m^{(\alpha,\,\beta)}(\rho)/d\rho=-2\pi\rho\Psi_{m+2}^{(\alpha,\,\beta)}(\rho)$ of the scaling function~\cite{Uchaikin:1999a}, one can easy find the spectral exponent $\eta$ for observed particles:
\begin{equation}\label{eq:eta}
\eta = -\frac{d\log N}{d\log E} = \gamma +\frac{\delta}{\alpha}\Xi,
\end{equation}
where
\begin{multline}\label{eq:xi}
\Xi = 3-\frac{2\pi r^2}{D(E,\alpha,\beta)^{2/\alpha}}\times\\
\times\frac{\int\limits_{\max[0,t-T]}^{t}d\tau \tau^{-5\beta/\alpha}\Psi_5^{(\alpha,\beta)}\left(|\rr|(D(E,\alpha,\beta)\tau^{\beta})^{-1/\alpha}\right)}
{\int\limits_{\max[0,t-T]}^{t}d\tau \tau^{-3\beta/\alpha}\Psi_3^{(\alpha,\beta)}\left(|\rr|(D(E,\alpha,\beta)\tau^{\beta})^{-1/\alpha}\right)}.
\end{multline}

Let $E_k$ be a solution of the equation $\Xi(E) = 0$. One can see from~\eqref{eq:eta} and~\eqref{eq:xi} that at $E=E_k$ the spectral exponent for observed particles $\eta$ is equal to the spectral exponent for particles
generated by the source: $\eta(E_k) = \gamma$. Since the exponent $\eta|_{E\ll E_k} = \gamma - \delta$ is less than $\gamma$ at $E\ll E_k$, but the exponent $\eta|_{E\gg E_k} = \gamma + \delta/\beta$ is greater than $\gamma$ at $E\gg E_k$, $E_k$ can be called the ``knee'' energy.

From experimental values of $\eta|_{E\ll E_k}$ and $\eta|_{E\gg E_k}$ one can derive the main parameters of the model $(\gamma,\delta)$ versus the spectral exponent $\beta$ of the ``L\'{e}vy waiting time'':
$$\delta = \left(\eta|_{E\gg E_k} - \eta|_{E\ll E_k}\right)\frac{\beta}{1+\beta},\quad \gamma = \eta|_{E\ll E_k} + \delta.$$

Since $\eta|_{E\gg E_k} - \eta|_{E\ll E_k}\sim 0.6$, $\eta|_{E\ll E_k}\sim 2.55\div 2.65$~\cite{Bartoli:2015}, $\delta\sim 0.27$, the last equations permit to retrieve self-consistently both spectral exponents $\gamma$ and $\beta$:
$$\gamma\sim 2.85\div 2.95,\quad \beta\sim 0.8.$$

We note that a similar steep spectrum of accelerated
particles has been observed from supernova remnants
W44~\cite{Abdo:2010} and IC 443~\cite{Abdo:2010_1} with the Fermi Large Area Telescope, and W49B with H.E.S.S. and Fermi-LAT~\cite{HESS:2016}. The value $\gamma = 3$ has been found in~\cite{Tanaka:2008} for RX J1713.7-3946. At energies higher than 400 GeV, VERITAS and Fermi-LAT observe gamma-ray emission from Tycho’s SNR with power-law index $\sim 2.92$~\cite{Archambault:2017}. 

To evaluate the parameter $\alpha$, general results for the particles' spectral exponent $\gamma$, obtained in the framework of the diffusive shock acceleration theory extended to the case of anomalous transport with ``L\'{e}vy flights'' and ``L\'{e}vy traps'',  have been used (see [Lagutin A.A., 2017, to be published]). It was shown that for $\beta=0.8$ the spectral exponent $\gamma=2.8\div 2.9$ corresponds to nondiffusive transport with $\alpha\sim 1.7$.

The technique for determining the anomalous diffusivity was described in~\cite{LagutinTymentcev:2004}. In this work, we used the following value $D_0\approx 1.5\cdot 10^{-3}$~pc$^{1.7}$y$^{-0.8}$.

As a result we have the cosmic ray spectrum

\begin{multline}\label{eq:sol}
	J(\rr,t,E) = \frac{v}{4\pi}\Biggl[S_{G}E^{-\gamma-\delta/\beta}+ \frac{S_0 E^{-\gamma}}{D(E,\alpha,\beta)^{3/\alpha}}\times\Biggr.\\
	 \times\sum\limits_{\substack{r_j < 1\;\text{kpc}\\t_j < 10^6\;\text{yr}}} \int\limits_{\max[0,t_j-T]}^{t_j}d\tau
	\tau^{-3\beta/\alpha}\Psi_3^{(\alpha,\beta)}\left(|\rr_j|(D(E,\alpha,\beta)\tau^{\beta})^{-1/\alpha}\right)+\\[-5mm]
	+\Biggl. S_{NS}\sum\limits_{r_j < 1\;\text{kpc}}E^{-\gamma+\delta_L}|\rr_j|^{-\alpha}\Biggr]\exp\left(-\frac{E}{E_0}\right).
      \end{multline}

A set of anomalous diffusion model parameters adopted in this paper is given in Table~\ref{tab:params}.

\begin{table}
\caption{Anomalous diffusion model parameters}\label{tab:params}
\begin{center}
\begin{tabular}{l|l}
\hline
Parameter & Value \\
\hline
$\alpha$ & $1.7$ \\
$\beta$ & $0.8$ \\
$D_0(\alpha,\beta)$ & $ 1.5\cdot 10^{-3}$~pc$^{1.7}$y$^{-0.8}$\\
$\delta$ & $0.27$ \\
$\delta_L$ & $\delta/2$ \\
$\gamma$ & $2.85$ \\
$E_0$ & $4\cdot 10^{18}Z$~eV \\
$T$ & $10^4$ y\\
\hline
\end{tabular}
\end{center}
\end{table}

\section{Results}

\begin{figure*}[ht!]
\centering
\includegraphics[width=0.9\textwidth]{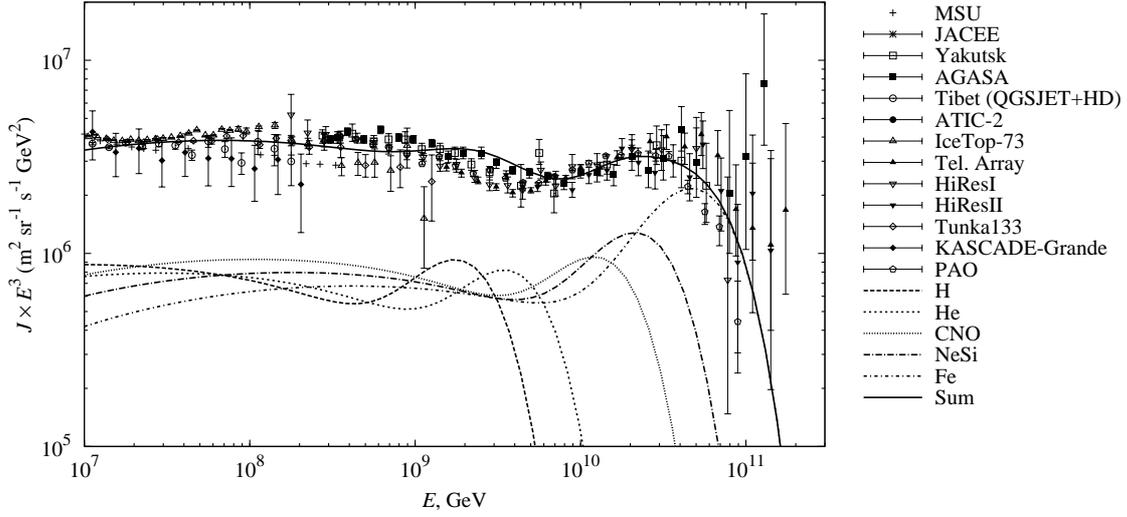}

\caption{All-particle spectrum obtained in the anomalous diffusion approach. Our results is compared with results from experiments: MSU~\cite{Fomin:1991}, JACEE~\cite{Asakimori:1998}, Yakutsk~\cite{Afanasiev:1995}, AGASA~\cite{Yoshida:1995}, Tibet~\cite{Amenomori:2008}, ATIC-2~\cite{Panov:2007,Panov:2009}, IceTop-73~\cite{Aartsen:2013}, Telescope Array~\cite{Abuzayyad:2015}, HiRes I~\cite{Abbasi:2004}, HiRes II~\cite{Abbasi:2008}, Tunka-133~\cite{Prosin:2014}, KASCADE-Grande~\cite{Apel:2013}, PAO~\cite{Aab:2013}}\label{fig:cr-spectrum}
\end{figure*}

In the framework of the model described above we have analysed the energy spectra and mass composition behaviour in the energy range up to the cut-off observed at $\sim 5\cdot 10^{19}$~eV. The energy spectra for all particles and elemental groups of nuclei are shown in Figure~\ref{fig:cr-spectrum}  compared with the data of different experiments. Along with good overall agreement with the representation of all the basic spectrum features, several key model predictions should be noted.

As  already mentioned, the anomalous diffusion provides the knee in the energy spectrum as a pure propagation effect without any specific assumptions about maximum energy that a particle can achieve in a source. Another important feature of the model is that at the energy of the knee the spectral exponent, $\eta$, is equal to the injection spectra exponent $\gamma$. Thus, if anomalous diffusion is, indeed, the main mechanism of the knee feature, then the injection spectra exponent could be estimated from the analysis of the experimentally measured energy spectra around the knee.

The energies of knee retrieved for five elemental groups of nuclei (H, He, CNO, NeMgSi, Fe) are shown in Table~\ref{Eknee}. One can see that the knee energies are almost proportional to $Z$. We note that our estimations for the H spectrum  are consistent with the results~\cite{Bartoli:2015} where the knee-like feature in the (p+He) spectrum is observed at energies around $700\pm 230$~TeV.

\begin{figure}%[ht!]
\centering

\includegraphics[width=.49\textwidth]{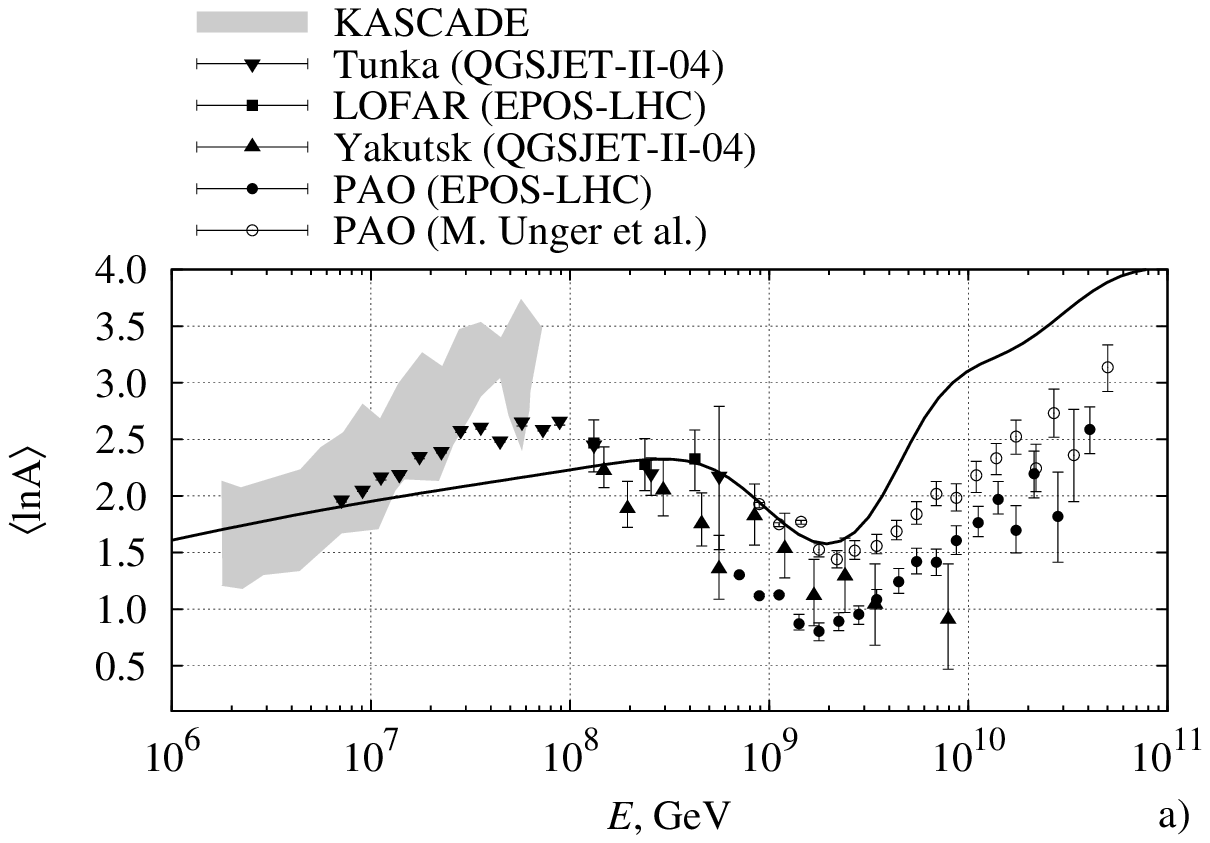}
\sidecaption
\includegraphics[width=.49\textwidth]{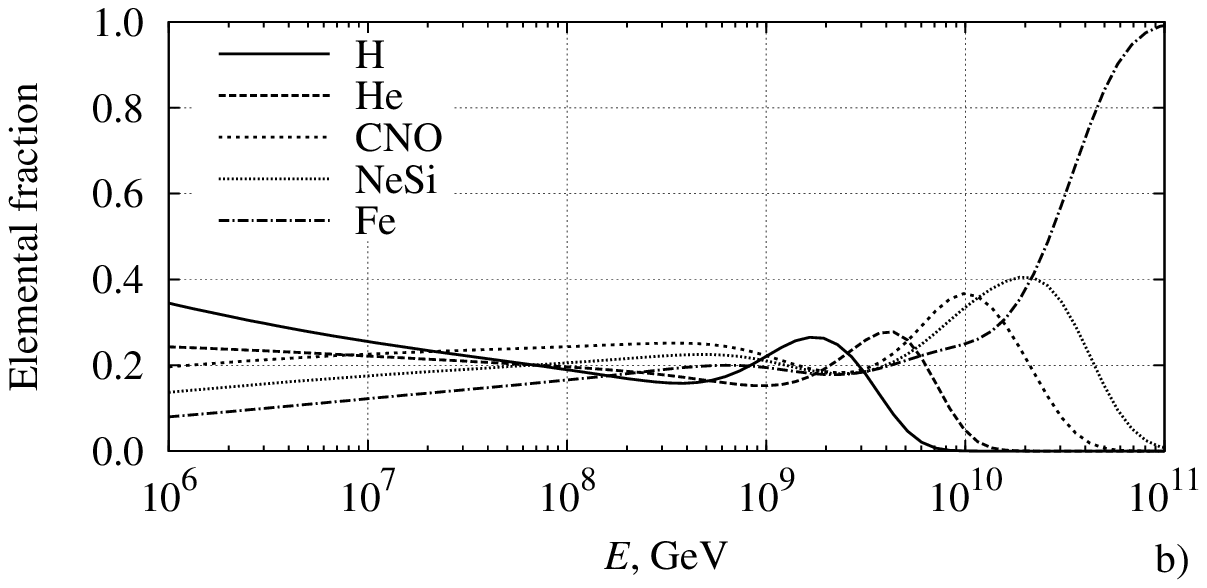}

\caption{Mean logarithmic mass (a) and elemental fractions (b) of CRs predicted by the anomalous diffusion model (lines). A shaded area and symbols show experimental results from~\cite{Thoudam:2016} (KASCADE, Tunka, LOFAR, Yakutsk), \cite{Aab:2014} (PAO) and \cite{Unger:2015} (PAO corrected by M.~Unger et al. for systematic uncertainties in energy and depth of maximum scales)}\label{fig:lna}
\end{figure}

Mean logarithmic mass and elemental fractions of CRs from $10^{15}$ to $10^{20}$~eV are shown in Figure~\ref{fig:lna}. According to our model, the mass composition became heavier with energy up to $\sim 4\cdot 10^{17}$~eV. It is seen that at $10^{17.5}$~eV our data is
consistent with LOFAR estimations~\cite{Buitink:2016}.

\begin{table}%[h!]
\caption{The knee energies for different elemental groups of nuclei}\label{Eknee}
\begin{center}
\begin{tabular}{l|l}
\hline
Nuclei & Knee energy, eV \\
\hline
H & $(6.1\div 6.5)\cdot 10^{14}$\\
He & $(1.8\div 2.2)\cdot 10^{15}$\\
CNO & $(5.8\div 6.3)\cdot 10^{15}$\\
NeMgSi & $(0.8\div 1.2)\cdot 10^{16}$\\
Fe & $(2.6\div 3.0)\cdot 10^{16}$\\
\hline
\end{tabular}
\end{center}
\end{table}

The non-scattered particles of atomic charge $Z$ contribute to the total flux starting from the energy $3\cdot 10^{17}Z$~eV, at which their gyroradius became large enough to escape from the typical Galactic inhomogeneity and reach the observer. This results in a decreasing mean mass of CRs in the energy region ($4\cdot 10^{17}\div 2\cdot 10^{18}$~eV) matching the PAO data~\cite{Settimo:2016} quite well. The lightest composition ($\langle\ln A\rangle \sim 1.6$) is observed at $\sim 2\cdot 10^{18}$~eV.

At ultra-high energies we have a gradually heavier composition with progressive cut-offs in each group of nuclei at $E_0=4\cdot 10^{18}Z$~eV. At $10^{19}$~eV our predictions give a heavier composition in comparison with the Auger data obtained from $X_{\max}$ distributions~\cite{Settimo:2016}, but agree qualitatively with recent estimations~\cite{Aab:2016} made on the basis of the advanced $X_{\max}$ and ground signal correlation analysis. 

Finally it should be noted that our model contains an ankle in the light component similar to that observed by KASCADE-Grande~\cite{Apel-a:2013} and more commonly the multiple ankle-like features in spectra of elemental groups of nuclei at $E>3\cdot 10^{17}Z$~eV. This should be considered as a crucial model prediction, which can be verified with the improved high-precision future measurements.

\section{Conclusions}

We have made a revision of the Galactic cosmic ray origin scenario in the framework of the anomalous diffusion model. It was shown that recent data on the  cosmic ray energy spectrum and mass composition can be well reproduced by our model over the entire primary energy range with a reasonable set of key assumptions.

The knee energies were retrieved for different elemental groups of nuclei showing the rigidity-dependent behaviour similar to the standard scenario prediction. It is shown that, if the knee feature is caused by the anomalous diffusion of particles through the fractal-like ISM, the source injection spectra exponent could be estimated from the experimentally observed energy spectra being equal to the spectral exponent at the knee energy.

The following basic model predictions are expressed.
\begin{itemize}
\item The suppression of the all-particle spectrum at \mbox{$E=5\cdot 10^{19}$~eV} is due to the nuclei fluxes cut-offs caused by the rigidity-dependent energy limitation of the Galactic sources.
\item The composition becomes heavier with energy from the knee to $\sim 10^{17.5}$~eV and reaches a maximum of mean logarithmic mass $\langle\ln A\rangle \sim 2.4$.
\item In the energy region ($4\cdot 10^{17}\div 2\cdot 10^{18}$~eV) the mean logarithmic mass decreases reaching the minimum value of $\langle\ln A\rangle \sim 1.6$.
\item The rapid weighting of the mass composition is observed at $E>2.5\cdot10^{19}$~eV up to the pure iron composition at the cut-off.
\item Multiple ankle-like spectrum features at energies above $3\cdot 10^{17}Z$~eV reflecting the contribution of the non-scattered CR component are expected.
\end{itemize}

\section*{Acknowledgements}
This work was supported in part by the Ministry of Education and Science of The Russian Federation (state assignment for the fundamental and applied research performed at  Altai State University).

\end{document}